\documentclass[
pre,reprint,longbibliography,onecolumn,
 amsmath,amssymb,
 aps,
]{revtex4-2}

\usepackage{bm}
\usepackage{graphicx}
\usepackage{enumitem}   
\usepackage{appendix}
\usepackage{soul}
\usepackage[hidelinks]{hyperref}
\usepackage{cleveref}
\usepackage{color}
\usepackage[normalem]{ulem}
\usepackage{comment}
\usepackage{subfig}

\begin{document}

\preprint{123-QED}

\title{Channel-facilitated transport under resetting dynamics}
\author{Suvam Pal}
\email{suvamjoy256@gmail.com}
\affiliation{Physics and Applied Mathematics Unit, Indian Statistical Institute, 203 B.T. Road Kolkata, India}
\author{Denis Boyer}
 \email{boyer@fisica.unam.mx}
 \affiliation{Instituto de F\'isica, Universidad Nacional Aut\'onoma de M\'exico, Ciudad de M\'exico C.P. 04510 M\'exico.}
\author{Leonardo Dagdug}
 \email{dll@xanum.uam.mx}
\affiliation{Physics Department, Universidad Aut\'onoma Metropolitana-Iztapalapa, San Rafael Atlixco 186, Ciudad de M\'exico, 09340, M\'exico.}
 \author{Arnab Pal}
 \thanks{Corresponding author}
 \email{arnabpal@imsc.res.in}
\affiliation{The Institute of Mathematical Sciences, CIT Campus, Taramani, Chennai 600113, India \& 
Homi Bhabha National Institute, Training School Complex, Anushakti Nagar, Mumbai 400094,
India}

\date{\today}

\begin{abstract}
The transport of particles through channels holds immense significance in physics, chemistry, and biological sciences. 
For instance, the motion of solutes through biological channels is facilitated by specialized proteins that create water-filled channels and valuable insights can be obtained by studying the transition paths of particles through a channel and gathering statistics on their lifetimes within the channel or their exit probabilities. In a similar vein, we consider a one-dimensional model of 
channel-facilitated transport where a diffusive particle is subject to attractive interactions with the walls within a limited region of the channel. We study the statistics of conditional and unconditional escape times, in the presence of resetting -- an intermittent dynamics that brings the particle back to its initial coordinate randomly. We determine analytically the physical conditions under which such resetting mechanism can become beneficial for faster escape of the particles from the channel thus enhancing the transport. Our theory has been verified with the aid of Brownian dynamics simulations for various interaction strengths and extent. The overall results presented herein highlight the scope of resetting-based strategies to be universally promising for complex transport processes of single or long molecules through biological membranes.
\end{abstract}

\maketitle

\section{Introduction}
The subject of diffusion in constrained geometries remains an enduring subject across diverse scales, encompassing biological and technological areas \cite{dagdugdiffusion}. For instance,  proteins responsible for forming water-filled channels inside membranes are pivotal in aiding the transport of metabolites and other solutes across cell membranes~\cite{rostovtseva2008tubulin,hoogerheide2017mechanism}. In addition to  practical applications, fundamental aspects of channel-regulated transport have been the focus of many theoretical explorations. See \cite{dagdugdiffusion} which provides an exhaustive list of these problems.

Diffusion processes in high dimensional confined geometries and their effective reduction to one-dimensional problems have been studied over nearly two centuries since the work of Fick, much later revisited by Jacobs and Zwanzig \cite{fick1855ueber,jacobs1967diffusion,zwanzig1992diffusion,berezhkovskii2011time}. Such a reduced description underlies a rapid equilibration of the faster degrees of freedom of the particles across the confinements, which is the key assumption behind the Fick-Jacobs formalism. Zwanzig studied unbiased diffusive motion in quasi-one-dimensional systems with varying geometric constraints. Depending on the constraints, he calculated an effective diffusivity \cite{zwanzig1992diffusion}, paving the way for further developments over the past decades \cite{dagdug2012projection}. It has been possible to identify key quantities controlling transport processes thanks to extensive studies on tubes and channels with simple geometries \cite{rubi2019entropic,haul1993j,han2000separation,gershow2007recapturing,dagdug2009drift,skvortsov2018trapping,verdel2016unbiased}. Despite numerous years of research, new phenomena linked to channel-assisted metabolite transport keep emerging, and transport in confinement remains of much topical interest.

Consider a particle that enters a channel of an arbitrary shape from one end. The particle can either exit on the same side of the membrane where it entered or it can pass through and exit from the other end \cite{berezhkovskii2003channel}. Essential observables to study transport properties of such particles are return and translocation probabilities, the lifetime of the particle inside the channel which can be computed from the escape time statistics of the particle from the channel. A myriad of research has been devoted to capturing the statistics of these key quantities over the years \cite{dagdug2003diffusion,berezhkovskii2006identity,alexander2017mean}. Furthermore, single-particle experiments have allowed for the direct observation of transition or first passage or exit trajectories from the channel \cite{chung2012single,berezhkovskii2020peculiarities,ferrer2024experimental}. The tracking of such trajectories allow us to get further insights on the time when a molecule crosses an activation barrier or two molecules undergo a chemical reaction \cite{dagdug2003diffusion,berezhkovskii2006identity,berezhkovskii2020peculiarities}. The average residence time of a diffusive particle inside a channel has been studied theoretically in different geometries ~\cite{reguera2006entropic, burada2007biased,burada2009diffusion} and remains the focal point of research in channel-facilitated problems.

Generically, first-passage processes play a pivotal role in diverse fields of chemistry, physics, and biology~\cite{redner2001guide,bray2013persistence,benichou2005optimal,zhang2016first,iyer2016first}. What is the time required to initiate a chemical reaction? Or how much time does it take for a stock market order to be fulfilled? First passage times are relevant to a number of phenomena and the expectations such as the mean first passage time can quantify the duration of a specific task. Conditional first passage observables, such as splitting probabilities and conditional first passage time densities, also play a key role for a deeper understanding of the physical problem when there are multiple pathways to complete a task.

Recently, the optimization of first passage processes via different protocols has attracted considerable attention and in particular, resetting mediated dynamics has been an extremely effective protocol to reduce the completion time of a complex search process~\cite{evans2011diffusion,evans2020stochastic,pal2017first,reuveni2014role,pal2023random}. Resetting usually works by interrupting an ongoing process at a certain rate and bringing it back to a pre-determined state, thus by confining the errand far away trajectories. This confining effect renders anomalous non-equilibrium behavior \cite{evans2013optimal,gupta2014fluctuating,majumdar2015dynamical,pal2015diffusion,mendez2016characterization,eule2016non} but then also improves the speed-up in search processes by curtailing trajectories which take longer time to complete the task \cite{pal2017first,evans2018run,pal2019first,pal2019landau,evans2020stochastic,bonomo2021first,pal2023thermodynamic,evans2011diffusion,ray2021mitigating,ghosh2023autonomous,ray2020diffusion,ray2020space,ray2021resetting,kumar2023universal,pal2023random}. The most remarkable paradigm is perhaps the diffusive transport for which the mean first passage time is usually diverging in a non-confining domain, however, it has been shown that resetting can make this time optimally finite ~\cite{evans2011diffusion}. Following this work, the efficiency of resetting processes has been studied rigorously in different research avenues and is found to have significant speed-up in random search processes spanning from physics, chemistry, biology, economics, computer science and many more ~\cite{biswas2023rate, sar2023resetting,domazetoski2020stochastic,singh2020resetting,jolakoski2023first,julian2024diffusion,mercado2021search,huang2021random,blumer2022stochastic,blumer2024combining}. The impact of resetting on first passage properties in simple one-dimensional confinement has also been extensively investigated in stochastic systems, featuring combinations of absorbing or reactive boundaries \cite{pal2019first,ahmad2019first,ahmad2022first,capala2023optimization}. Recent experiments on resetting have also showcased these features and moreover, they have posed many interesting and new research directions \cite{tal2020experimental,besga2020optimal,faisant2021optimal,goerlich2023experimental,altshuler2024environmental,paramanick2024programming,paramanick2024prx}. 

However, the effect of resetting in a diffusion mediated \textit{channel transport} is the subject matter of current study and it was only recently Jain \textit{et al} 
addressed the transport of diffusing particles under resetting inside a three-dimensional conical channel for the first time \cite{jain2023fick}. There, it was shown that 
the interplay between resetting rate and the effective potential
created by the position dependent radius of the channel can give rise to a plethora of rich behavior for the optimal transport. Motivated by this work, in this study, we investigate the transport of resetting Brownian particles inside a membrane whose geometry can be visualized as a cylindrical channel with attractive interactions between the channel and the particle \cite{berezhkovskii2003channel}. The motion of the particles can then be described by the Fick-Jacobs formalism \cite{berezhkovskii2003channel,dagdugdiffusion}. Using this, we examine the dependence of first passage statistics on the particle/channel interaction strength and on the resetting rate. We also compute the escape probabilities and times of escape conditioned on the boundary. It is observed that resetting can indeed expedite transport of particles through membrane channel -- we find the corresponding phase space spanned by the parameters such as the potential strength and initial configuration of the particles where resetting can be beneficial, otherwise not. We also study in detail the behavior of the resetting rate whence it makes the transport optimal or the most efficient. We perform extensive Brownian dynamics simulations that corroborate with our theoretical observations.


\begin{figure}
    \centering
    \includegraphics[width=0.6\columnwidth]{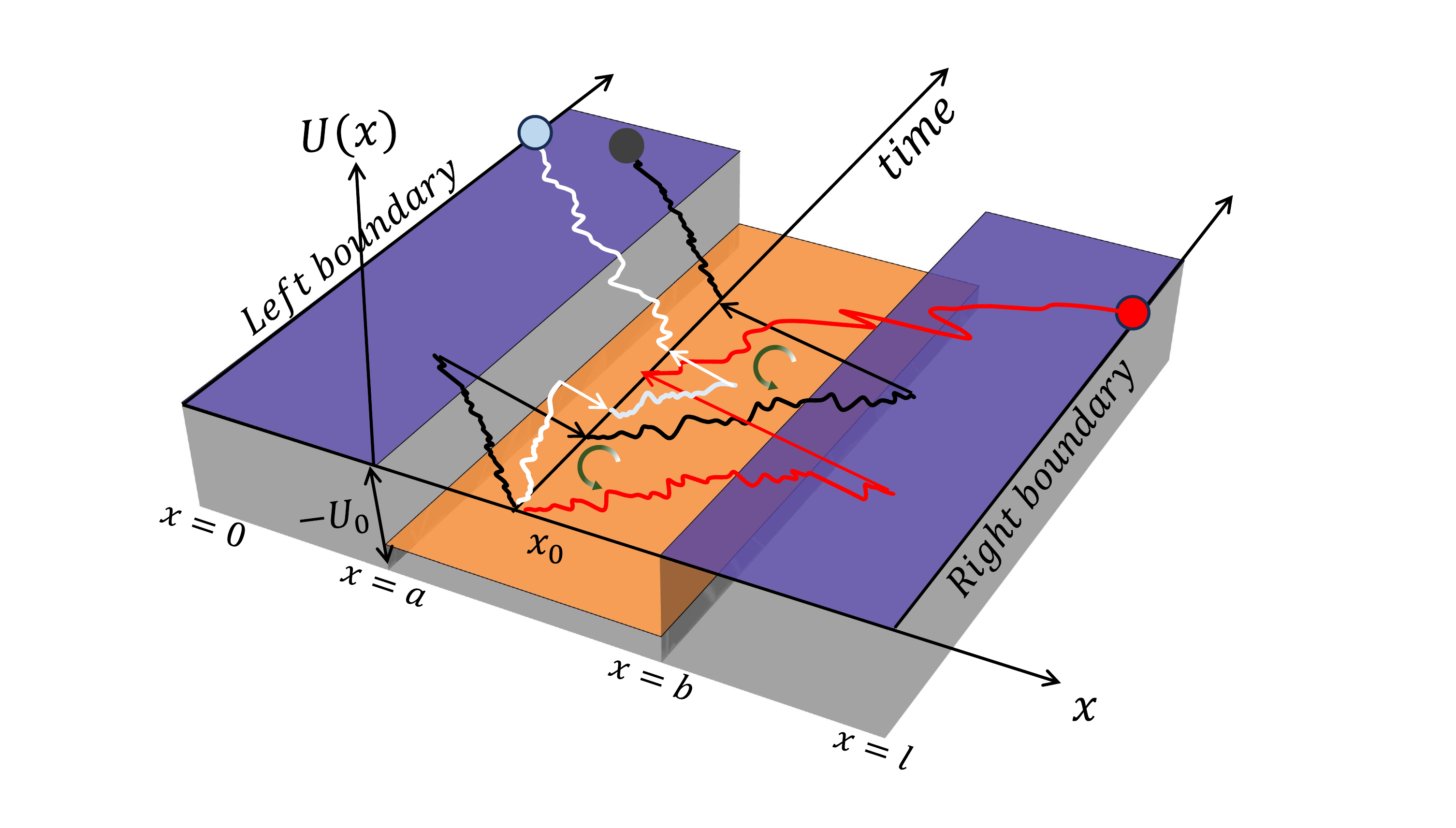}
    \caption{Channel-facilitated transport under resetting: Schematic of diffusing trajectories in a channel containing a potential of width $-U_0$.  The potential is spanned for a distance $b-a$. The entire channel is confined within two boundaries, left at $x=0$ and right at $x=l$, with a total distance of $l$.  The trajectories initiate from $x_0$ and are also reset to $x_0$ intermittently. The paths (in black) represent unconditional trajectories that escape from any of the boundaries. The trajectories in white and red represent the paths conditioned on the exit to the left or right boundary, respectively. Each trajectory gives rise to different escape time statistics which we study here.}
    \label{fig-1}
\end{figure}

The rest of the paper is organized in the following way. In Sec.~\ref{model-wo-reset}, we discuss the model and the solution of the Smoluchowski equation in the presence of a piece-wise constant potential. We derive matching conditions for the probability density function and the survival probability at the potential discontinuities. Also, we discuss the moments of the first passage time density related to the survival probability. In Sec.~\ref{model-w-reset}, we expose the formalism of the dynamics in the presence of resetting. In Subsection.~\ref{fj-surv-w-reset}, we study the variations of the mean first passage time as a function of the resetting rate and initial location of the particle. The investigation of the effect of resetting is discussed in Subsection.~\ref{cv}, followed by the optimal resetting rate is discussed in Subsection.~\ref{opt-r}. In Sec.~\ref{fj-pdf-w-reset} we elaborate on the conditional mean first passage times and the splitting probabilities for various initial locations and also portray how these quantities change qualitatively as the resetting rate is varied. We conclude in Sec.~\ref{conclusion}.

\section{Model details, formalism and quantities of interest}\label{model-wo-reset}
\subsection{Fick-Jacobs equation}\label{fj-pdf-wo-reset}
Let us consider the diffusive motion of a particle inside a cylindrical channel of length $l$ (see Fig. \ref{fig-1}). Our analysis is based on the Fick-Jacobs \& Zwanzig formalism where we approximate the 3D motion of a particle inside the channel as a one-dimensional diffusion along the channel axis \cite{zwanzig1992diffusion,dagdugdiffusion}. The applicability of this approach to three-dimensional cylinders crucially depends on the assumption of faster equilibration in the transverse degrees of freedom than in the longitudinal direction (\textit{i.e.} along the $x$-axis) \cite{zwanzig1992diffusion}. The model also assumes an interaction of the particle with the walls of the channel, in terms of a potential $U(x)$ (see Fig.~\ref{fig-1}). The fundamental observable in this problem is the density of the position $x$ at time $t$ for a particle starting at $x_0$, which can be denoted as $p_0(x,t|x_0,0)$ [for simplicity, we have used $p_0(x,t)$]. The 
evolution of the probability density function is given by the Smoluchowski equation \cite{dagdugdiffusion}
\begin{equation}\label{pdf}
    \dfrac{\partial}{\partial t} p_0\left(x,t \right)=D\dfrac{\partial}{\partial x}\left[e^{-\beta U(x)}\dfrac{\partial}{\partial x}\left(e^{\beta U(x)}p_0\left(x,t \right)\right)\right],
\end{equation}
with absorbing boundary conditions at the two ends of the interval
\begin{align}\label{bc-pdf}
    p_0\left(0,t\right)=0~,~~~p_0\left(l,t\right)=0.
\end{align}
 The initial condition is $p_0(x,0)=\delta\left(x-x_0\right)$, where $\beta=\dfrac{1}{k_BT}$ is the inverse temperature and is related to the diffusion constant by the fluctuation-dissipation relation namely $\beta=\mu/D$ where $\mu$ is the mobility of the particle. The form of the effective potential that the particle experiences in the channel can be considered as a potential well as described in \cite{berezhkovskii2003channel} and is given below  (also see Fig.~\ref{fig-1})
\begin{equation}\label{potential-form}
    U(x)=
    \begin{cases}
        0 & 0\leq x \leq a\\
        -U_0 & a< x< b\\
        0 & b\leq x\leq l,
    \end{cases}
\end{equation}
with $U_0>0$ and $0<a<b<l$. The nature of the potential 
suggests that Eq. (\ref{pdf}) can be solved in three different regions and then complemented with suitable matching conditions at the edges and boundary conditions at the boundaries. The governing equations in each region read
\begin{eqnarray}\label{piece-wise-pdf}
    \dfrac{\partial }{\partial t}p_0^L(x,t)=D\dfrac{\partial^2}{\partial x^2}p_0^L(x,t)~~~~~~~~~~0< x< a,\nonumber\\
    \dfrac{\partial }{\partial t}p_0^M(x,t)=D\dfrac{\partial^2}{\partial x^2}p_0^M(x,t)~~~~~~~~~~a< x< b,\\
    \dfrac{\partial }{\partial t}p_0^R(x,t)=D\dfrac{\partial^2}{\partial x^2}p_0^R(x,t)~~~~~~~~~~b< x< l,\nonumber
\end{eqnarray}
where $p_0^{L,M,R}(x,t)$ stands for the solutions in the left, middle, and right segments of the channel, respectively.

It is important to note that the potential has sharp jumps at $x=a$ and $x=b$ and thus the matching conditions have to be supplemented appropriately. We note that the fluxes are continuous across the jump coordinates, however the probabilities are discontinuous at these points due to the jump in potential. These are known as the \textit{imperfect} boundary conditions in the literature \cite{carr2016semi,farago2021thermodynamics,sheils2017multilayer,korabel2011boundary} and can be written for our set-up in the following way
\begin{equation}\label{cont-bc}
    \begin{cases}
          \partial_x p_0^M\left(x=a,t\right)=\partial_x p_0^L\left(x=a,t\right),\\
          \partial_x p_0^R\left(x=b,t\right)=\partial_x p_0^M\left(x=b,t\right).
    \end{cases}
\end{equation}
and
\begin{equation}\label{jump-bc}
    \begin{cases}
        p_0^L\left(x=a,t\right)=e^{-\beta U_0}p_0^M\left(x=a,t\right),\\
        p_0^R\left(x=b,t\right)=e^{-\beta U_0}p_0^M\left(x=b,t\right).
    \end{cases}
\end{equation}

Using the boundary conditions \eqref{bc-pdf} and the matching conditions \eqref{cont-bc} and \eqref{jump-bc}, the probability density functions can be solved exactly. For brevity, we have relegated these details to the Appendix~\ref{fw-approach} and we will recall the solutions wherever needed in context.

\begin{figure}
    \centering
    \includegraphics[width=0.5\columnwidth]{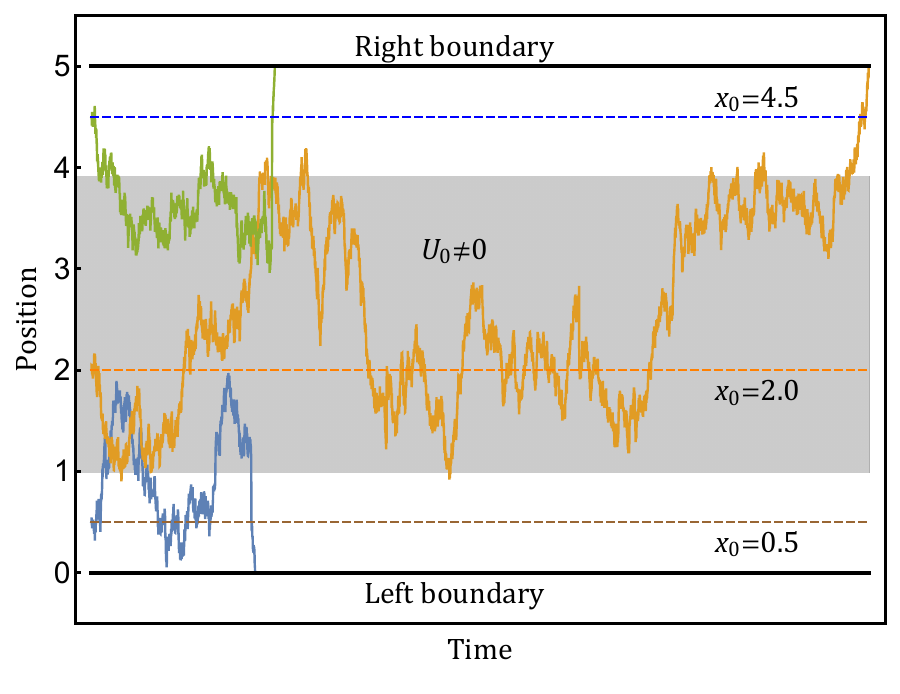}
    \caption{Schematic representation of trajectories for different initial locations:  $x_0=0.5$ (blue), $x_0=2$ (orange), and $x_0=4.5$ (green) in the absence of resetting. The shaded area represents regions with non-zero potential strength, while it is zero in the rest of the channel.}
    \label{fig-2}
\end{figure}

\subsection{Survival probability and first passage}\label{fj-surv-wo-reset}
Since we are interested into the lifetime of the particles inside the channel, a useful quantity to measure the same is the mean first passage time. However, to compute the mean first passage time one needs to have the knowledge of the first passage time density which can be estimated from the survival probability. In this section, we provide a brief outline of this formalism. Formally, the survival probability $Q_0(t|x_0)$ is defined as the probability that the particle is inside a prescribed domain without getting absorbed upto time $t$ given that it had started from an initial position $x_0$. Thus, $Q_0(t|x_0)=\int_{0}^{l}p_0(x,t|x_0,0)~dx$. Although the survival probability can be computed from the probability density function, there is an alternative route via backward path formalism that allows one to directly write a governing equation for the survival probability without an explicit knowledge of the probability density function. This is given by the backward Smoluchowski equation where the initial location $x_0$ is considered as a variable 
\begin{equation}\label{fp-wo-res}
    \frac{\partial}{\partial t}Q_0\left(t|x_0\right)=D e^{ \beta U(x_0)}\frac{\partial}{\partial x_0}\left[e^{-\beta U(x_0)}\frac{\partial}{\partial x_0}Q_0\left(t|x_0\right)\right].
\end{equation}
The above equation is supplemented with the boundary conditions
 $Q_0\left(t|x_0=0\right)=0$, $Q_0\left(t|x_0=l\right)=0$ and an initial condition $Q_0\left(0|x_0\right)=1$ for $0<x_0<l$. As mentioned before, due to the nature of the potential, we can solve the survival probability region-wise and then complement with the appropriate matching conditions at the jump coordinates. Translating Eq.~\eqref{fp-wo-res} into the Laplace space, we can write 
\begin{eqnarray}
    s\widetilde{Q}_0^I\left(s|x_0\right)-1=D\frac{\partial^2}{\partial x_0^2}\widetilde{Q}_0^I\left(s|x_0\right)~~~~0< x_0< a\label{fp-eqs-wo-res-lap,1},\\
    s\widetilde{Q}_0^{II}\left(s|x_0\right)-1=D\frac{\partial^2}{\partial x_0^2}\widetilde{Q}_0^{II}\left(s|x_0\right)~~~~a< x_0< b\label{fp-eqs-wo-res-lap,2},\\
    s\widetilde{Q}_0^{III}\left(s|x_0\right)-1=D\frac{\partial^2}{\partial x_0^2}\widetilde{Q}_0^{III}\left(s|x_0\right)~~~~b< x_0< l,\label{fp-eqs-wo-res-lap,3}
\end{eqnarray}
where $\widetilde{Q}_0^{I,II,III}(s|x_0)=\int_0^{\infty}dt\, e^{-st}Q_0^{I,II,III}(t|x_0)$ denote the solutions in the left, middle and right intervals, respectively in the Laplace space. 
The matching conditions for the survival probabilities at the jump coordinates read (detailed derivation of these matching conditions are provided in Appendix-\ref{appendix-b}) 
\begin{equation}\label{imperfect-bound}
    \begin{cases}
        \partial_{x_0} \widetilde{Q}_{0}^{I}\left(s|x_0=a\right)=e^{\mu U_0/D}\partial_{x_0} \widetilde{Q}_{0}^{II}\left(s|x_0=a\right),\\
        \partial_{x_0} \widetilde{Q}_{0}^{III}\left(s|x_0=b\right)=e^{\mu U_0/D}\partial_{x_0} \widetilde{Q}_{0}^{II}\left(s|x_0=b\right),
    \end{cases}
\end{equation}
and 
\begin{equation}\label{flux-cont}
    \begin{cases}
    \widetilde{Q}_{0}^{I}\left(s|x_0=a\right)=\widetilde{Q}_{0}^{II}\left(s|x_0=a\right),\\
    \widetilde{Q}_{0}^{III}\left(s|x_0=b\right)=\widetilde{Q}_{0}^{II}\left(s|x_0=b\right).
    \end{cases}
\end{equation}

Solving Eqs.~\eqref{fp-eqs-wo-res-lap,1}-\eqref{fp-eqs-wo-res-lap,3} and imposing the boundary conditions at $x=0$ and $x=l$, the quantities $Q_0^I$, $Q_0^{II}$ and $Q_0^{III}$ take the form
\begin{align}\label{sols-wo-res}
    &\widetilde{Q}_0^{I}\left(s|x_0\right)=\frac{1}{s}(1-e^{-\alpha_0 x_0})\left[1+A\,s\left(e^{\alpha_0x_0}+1\right)\right],\nonumber\\
    &\widetilde{Q}_0^{II}\left(s|x_0\right)=\frac{1}{s}+C_1e^{\alpha_0x_0}+C_2e^{-\alpha_0x_0},\\
    &\widetilde{Q}_0^{III}\left(s|x_0\right)=\frac{1}{s}[1-e^{-\alpha_0(x_0-l)}]\left[1+B\,s\left(e^{\alpha_0x_0}+e^{\alpha_0l}\right)\right]\nonumber,
\end{align}
where $\alpha_0=\sqrt{s/D}$. The constants $A, B, C_1$ and $C_2$ are obtained using the four matching conditions in Eqs.~\eqref{imperfect-bound}~and~\eqref{flux-cont}. Details are provided in Appendix-\ref{relation-coeff}. Using Eqs.~\eqref{sols-wo-res}, one can now write the complete solution for the survival probability 
\begin{equation}\label{bw-surv}
    \widetilde{Q}_0\left(s|x_0\right)=\Theta\left(a-x_0\right)\widetilde{Q}^I_0\left(s|x_0\right)+\Theta\left(x_0-a\right)\Theta\left(b-x_0\right)\widetilde{Q}^{II}_0\left(s|x_0\right)+\Theta\left(x_0-b\right)\widetilde{Q}^{III}_0\left(s|x_0\right).
\end{equation}

The first passage time density $f_0(t)$ is related to the rate of change of the survival probability so that $f_0(t)=-dQ_0(t)/dt$. Subsequently, one can compute all the moments from the FPT density directly
\begin{equation}\label{tn}
    \langle T^n\rangle=\int_{0}^{\infty}t^n~f_0(t)~dt,
\end{equation}
In particular, the mean first passage time can be obtained by setting $n=1$ and after some simplification, one finds
\begin{align}
    \langle T\rangle=\int_{0}^{\infty}t~f_0(t)~dt=\widetilde{Q}_0(s\rightarrow 0|x_0),
\end{align}
where $\widetilde{Q}_0\left(s|x_0\right)$ is given by Eq. (\ref{bw-surv}). Thus, from the knowledge of the survival function in the Laplace space, one can easily compute the mean lifetime of a particle inside a given geometry.

\section{First passage properties under Resetting dynamics}\label{model-w-reset}
So far, we have focused on the diffusive transport of the particle through the channel. In what follows, we will introduce the resetting mechanism to the original dynamics and study its effects on the first passage statistics. Resetting is a simple mechanism by which motion of a particle is intermittently stopped and it is reset back to its initial coordinate. Following this action, the particle restarts its motion and continues to diffuse till the next resetting occurs. The overall search process ends when the particle escapes from one of the exit points of the channel. We assume that resetting occurs at random times and that they are distributed according to $f_R(t)=re^{-rt}$, where $r$ is the rate at which resetting occurs. Thus the mean waiting time between two resetting events is given by $1/r$.



\subsection{Survival probability under resetting and unconditional MFPT}\label{fj-surv-w-reset}
We start by defining the survival probability $Q_r\left(t|x_0\right)$ under the resetting dynamics. Formally, it can be interpreted as the probability that the particle remains inside the channel upto time $t$ in the presence of repeated resetting to the initial position $x_0$. Using the renewal techniques \cite{evans2020stochastic,pal2019first}, one can write
    \begin{equation}\label{last-renew}
        Q_r\left(t|x_0\right)=e^{-rt}Q_0\left(t|x_0\right)+r\int_{0}^{t} d\tau e^{-r\tau}Q_r\left(t-\tau|x_0\right)Q_0\left(\tau|x_0\right),
    \end{equation}
where the first term on the right-hand side (rhs) of Eq.~\eqref{last-renew} implies that the particle survives until time $t$ and that no resetting event has occurred, whereas the second term of the rhs assumes possibility of one or multiple resetting events in the interval $[0,t]$. Here, we have used the fact that the probability of not having any resets upto time $t$ is given by $e^{-rt}$ and the probability of not exiting upto time $t$ without resetting is given by $Q_0(t|x_0)$. 
Taking Laplace transformation on the both sides of Eq.~\eqref{last-renew} 
and solving for $\widetilde{Q}_r$, we obtain
\begin{equation}\label{last-renew-lap}
    \widetilde{Q}_r\left(s|x_0\right)=\dfrac{\widetilde{Q}_0\left(r+s|x_0\right)}{1-r\widetilde{Q}_0\left(r+s|x_0\right)},
\end{equation}
which connects the survival probabilities for the processes with and without resetting. Utilizing Eq. (\ref{bw-surv}) into Eq. 
\eqref{last-renew-lap}, we find
\begin{equation}\label{bw-surv-reset}
    \widetilde{Q}_r\left(s|x_0\right)=\dfrac{\Theta\left(a-x_0\right)\widetilde{Q}^I_0\left(r+s|x_0\right)+\Theta\left(x_0-a\right)\Theta\left(b-x_0\right)\widetilde{Q}^{II}_0\left(r+s|x_0\right)+\Theta\left(x_0-b\right)\widetilde{Q}^{III}_0\left(r+s|x_0\right)}{1-r\left(\Theta\left(a-x_0\right)\widetilde{Q}^I_0\left(r+s|x_0\right)+\Theta\left(x_0-a\right)\Theta\left(b-x_0\right)\widetilde{Q}^{II}_0\left(r+s|x_0\right)+\Theta\left(x_0-b\right)\widetilde{Q}^{III}_0\left(r+s|x_0\right)\right)}, 
\end{equation}
where the region-wise survival probabilities $\widetilde{Q}^I_0,~\widetilde{Q}^{II}_0$ and $\widetilde{Q}^{III}_0$ are given by Eqs.~\eqref{sols-wo-res}. The mean first passage time (MFPT) from the channel, denoted by $\langle T_r\left(x_0\right)\rangle$, in the presence of resetting
can be obtained as
\begin{equation}\label{mfpt}
    \langle T_r\left(x_0\right)\rangle=\widetilde{Q}_r\left(s\rightarrow 0|x_0\right).
\end{equation}

\begin{figure}
    \centering
    \includegraphics[width=0.6\columnwidth]{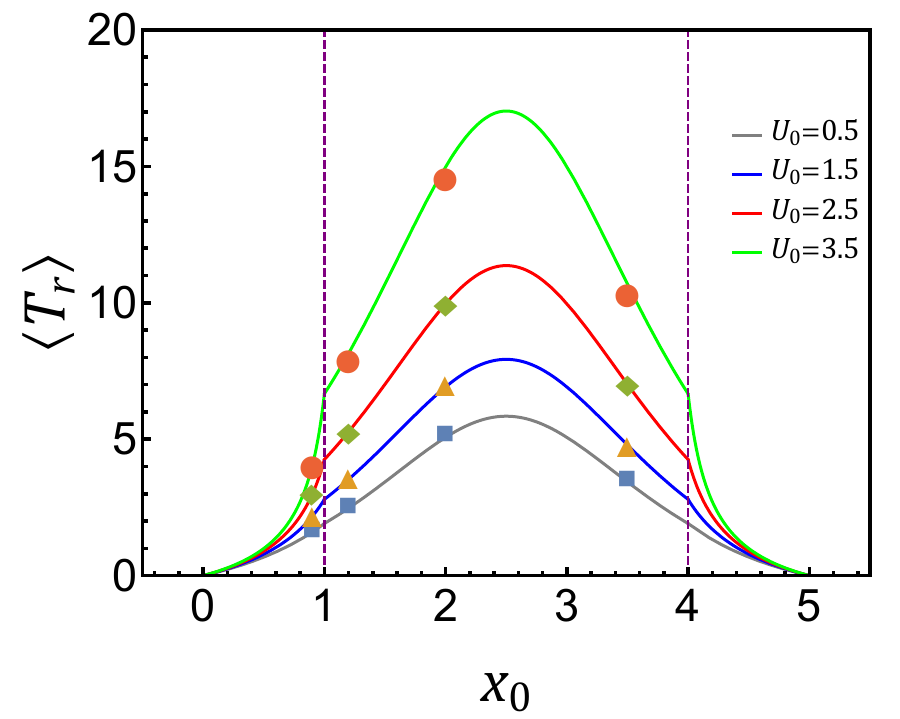}
    \caption{Unconditional mean first-passage times $\langle T_r\rangle$ vs. the initial position $x_0$ for a fixed value of the resetting rate $r=1$ and several potential strengths. The solid lines depict the theoretical results derived from Eq.~\eqref{mfpt}, whereas the markers indicate the values obtained from numerical simulations. We have set $D=1,~\mu=0.5$, $l=5$. The potential sink is comprised in between the two vertical lines, at $a=1$ and $b=4$. The cusps in this figure emanate from the matching conditions Eqs.~\eqref{imperfect-bound} and \eqref{flux-cont}, which demonstrate that although the survival propagator remains continuous across the potential boundaries, their first derivatives are discontinuous due to the inclusion of a pre-factor $e^{\mu U_0/D}$. This results in the cusps at the potential boundaries/edges.}
        \label{fig-3}
\end{figure}

In Fig.~\ref{fig-3}, we showcase the behavior of $\langle T_r\left(x_0\right)\rangle$ as a function of the initial position $x_0$ for a fixed resetting rate. We note that the MFPT increases with the potential depth $U_0$, but this effect is milder if $x_0$ is close to one of the boundary. We find a good agreement between theoretical and numerical results. The simulation method has been outlined in Appendix \ref{simul}. 
In Fig.~\ref{fig-4}, we illustrate the behavior of $\langle T_r\left(x_0\right)\rangle$ with respect to the resetting rate for three initial positions. We observe that resetting can both accelerate or prolong the escape time of the particle from the channel. For instance, when the particle starts close to the left or right boundary, i.e., $x_0=0.9$ (orange line) or $x_0=4.5$ (green line), we find that resetting reduces $\langle T_r\left(x_0\right)\rangle$ and in fact, it even attains a minima for some resetting rate. For an initial position of $x_0=2$, i.e., when the particle starts inside the potential sink, $\langle T_r(x_0)\rangle$ turns out to increase monotonously with $r$, hence resetting delays the search time or increases the lifetime of the particle. Thus, depending on the initial location and potential shape, the resetting strategy is found to have adverse effect on the particle exit from the channel. It is therefore natural to ask whether there is a general principle that governs the effect of resetting. We elaborate on this in the next subsection.

\begin{figure}[ht!]
    \centering
    \includegraphics[width=0.6\columnwidth]{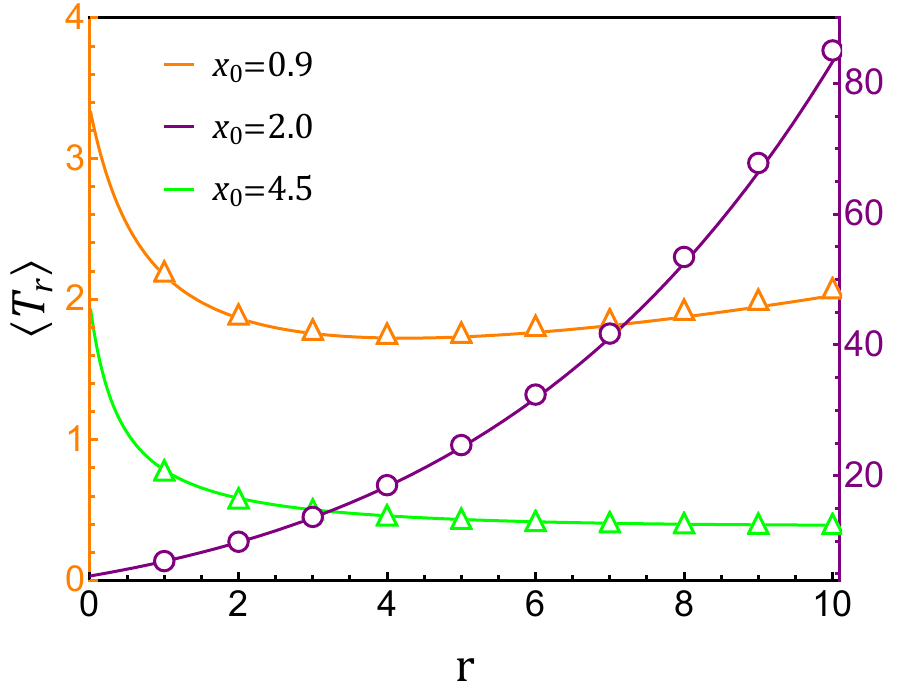}
    \caption{In this figure, we represent the unconditional MFPT as a function of $r$ for three initial positions $x_0=0.9,~2.0$, and $4.5$ as mentioned in Eq.~\eqref{mfpt}. Solid lines correspond to the theoretical estimation from Eq.~\eqref{mfpt} and the symbols represent data points from the numerical simulations. Here, we have set: $U_0=1.5,~a=1,~b=4$, $D=1,~\mu=0.5$, and $l=5$.}
    \label{fig-4}
\end{figure}

\subsection{$CV$ criterion}\label{cv}
\begin{figure}
    \centering
    \includegraphics[width=0.6\textwidth]{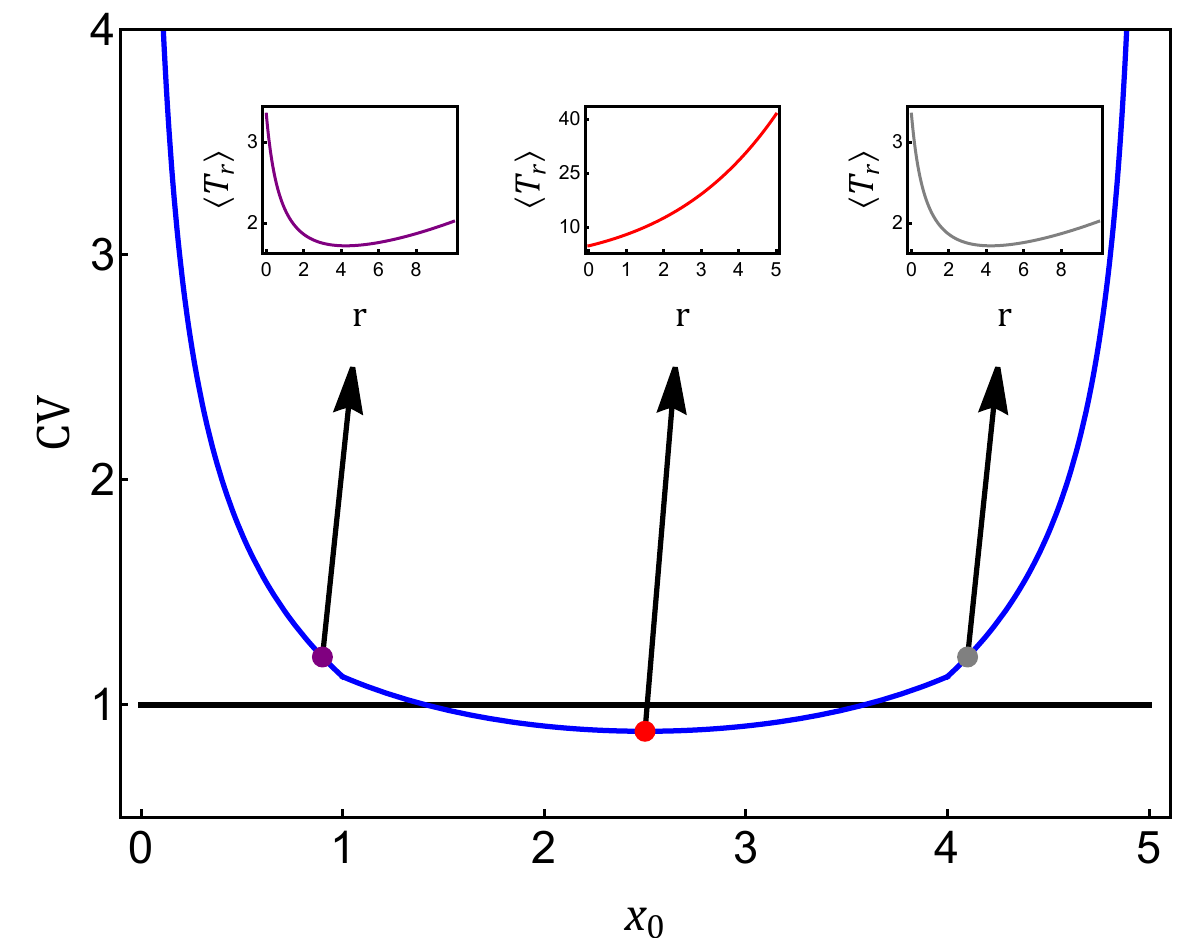}
    \caption{Coefficient of variation $CV$ in terms of the initial position $x_0$, for an symmetric potential with spread $a=1$, $b=4$ and depth $U_0=1.5$ in a channel of length $l=5$. The horizontal dashed line corresponds to $CV=1$ and it separates the region above and below $CV>1$ as mentioned in Eq.~\eqref{cv-criteria}.  The mean first passage time $\langle T_r \rangle$ as a function of the resetting rate is also shown for three different $x_0$ with different $CV$. For this figure, we use the same set of parameters as mentioned in Fig.~\ref{fig-4}.}
    \label{fig-5}
\end{figure}

\begin{figure}[ht!]
    \centering
    \subfloat[]{\includegraphics[width=0.6\columnwidth]{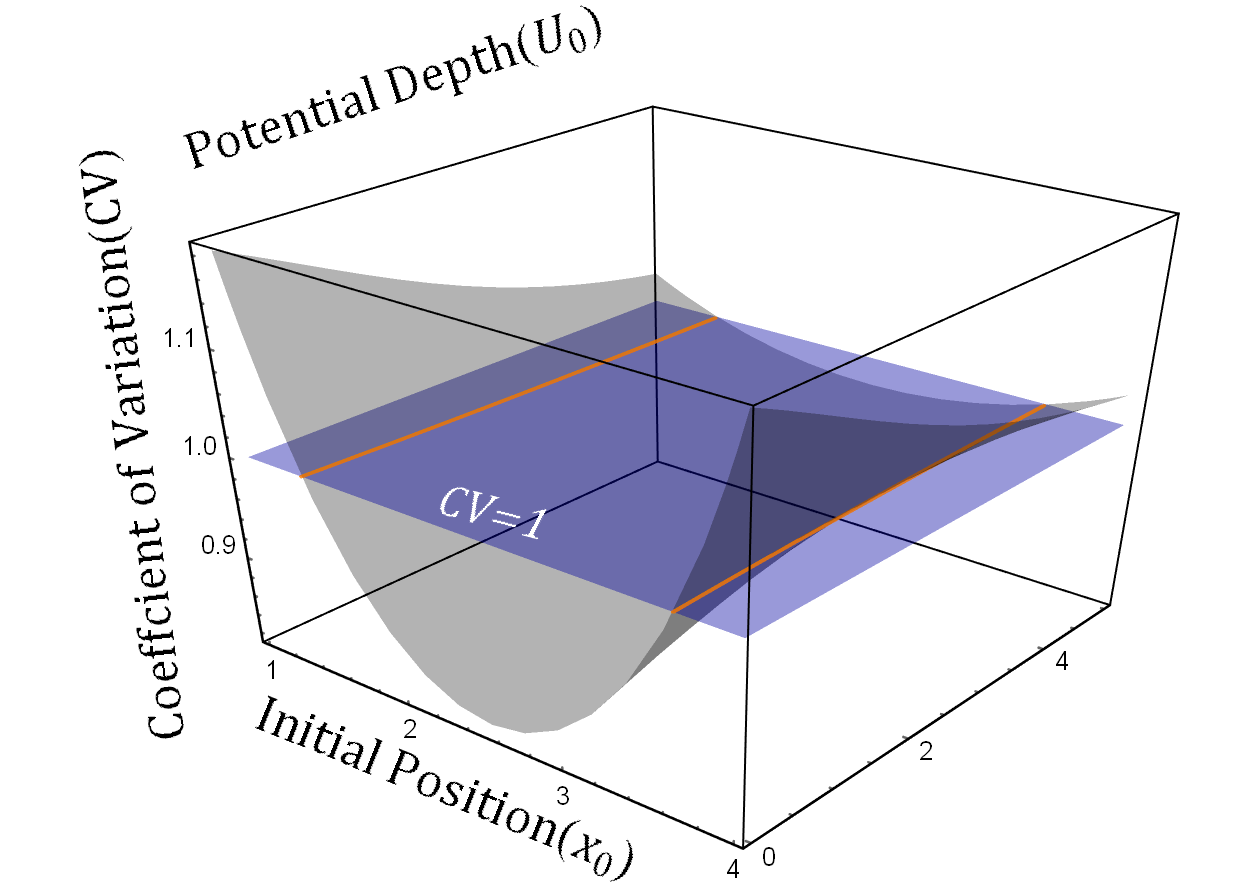}}
    \subfloat[]{\includegraphics[width=0.4\columnwidth]{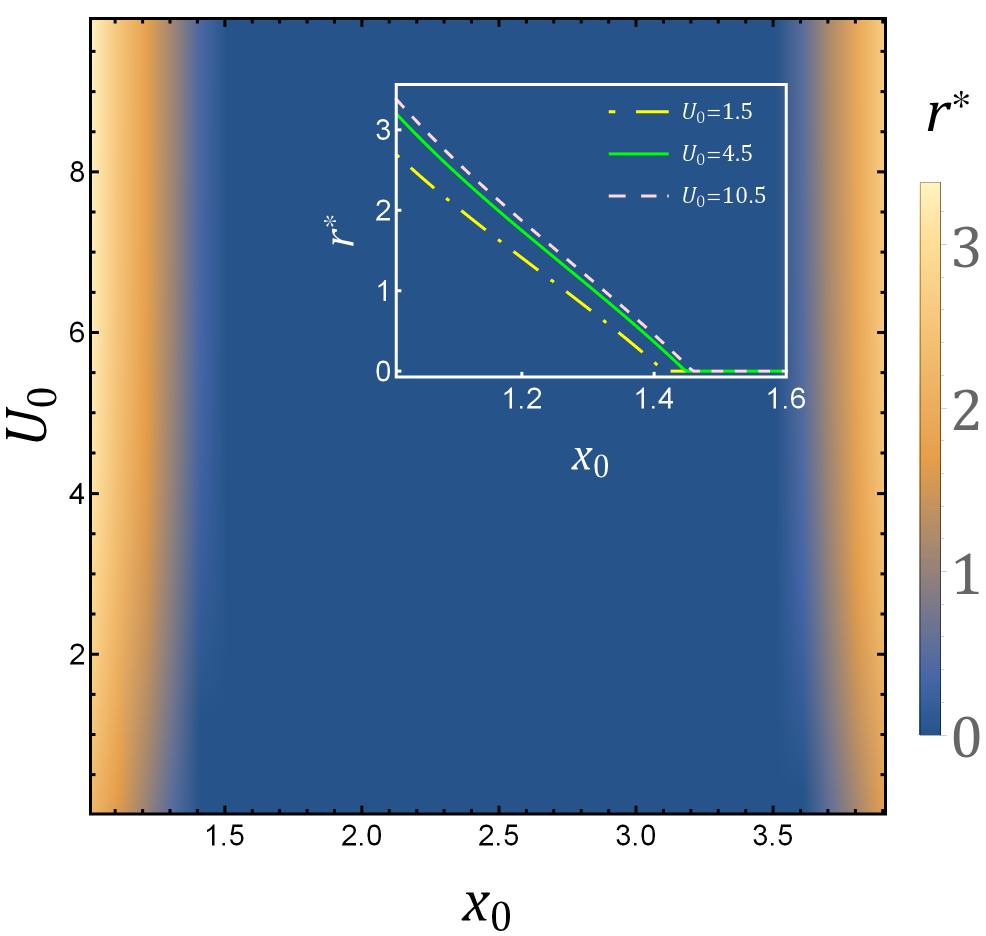}}
    \caption{Panel (a):  Variation of $CV$ in the phase space spanned by the initial position $x_0$ and potential depth $U_0$. The horizontal plane indicates $CV=1$, as mentioned in Eq.~\eqref{cv-criteria} and here, the orange lines indicate the intersection of two surfaces. The other parameters are $a=1$, $b=4$ and $l=5$. Panel (b): Variation of the optimal resetting rate $r^*$, as defined in Eq.~\eqref{opt-r-eq} with respect to $x_0$ and $U_0$. The color bar indicates the value of $r^*$ across that phase space. 
    The inset in panel (b) depicts the variation of the optimal resetting rate $r^*$ in terms of the initial position $x_0$ for different $U_0$. It is seen that $r^*$ continuously undergoes a transition from a non-zero value to zero thus marking a `resetting transition' as explained in the main text.} 
    \label{fig-6}
\end{figure}

It is now well understood that if the resetting free first passage statistics have large fluctuations in search time, then resetting is guaranteed to expedite the search \cite{pal2017first,pal2023random,pal2022inspection}. This is quantified by the so-called $CV$-condition 
\begin{equation}\label{cv-criteria}
    CV>1,
\end{equation}
where $CV$ is the coefficient of variation and its squared value is defined as $CV^2=\left[\langle T_0(x_0)^2\rangle-\langle T_0(x_0)\rangle^2\right]/\langle T_0(x_0)\rangle^2$ -- a dispersion ratio between the standard deviation and the mean of the first passage time in the absence of resetting. In Fig.~\ref{fig-5}, for a given potential configuration, we have plotted $CV$ as a function of the initial condition $x_0$ alongside the phase boundary $CV=1$ which is represented by the dashed horizontal line. This line separates between the resetting -beneficial and -detrimental regions spanned by $x_0$. It is seen that resetting is useful near the edges while it remains detrimental through out the bulk of the channel. The insets show the behaviors of the MFPT with $r$ for three different initial conditions corroborating with our observations. Extending this observation, we now plot $CV$ (Fig.~\ref{fig-6}(a)) as a function of $x_0$ for a symmetric potential well ($a=1$, $b=4$ and $l=5$) for different potential depth $U_0$ alongside the phase boundary $CV=1$. It can be noticed that as $U_0$ increases, the range of values of $x_0$ for which resetting is beneficial widens slightly. This implies that in the presence of an attractive potential, resetting represents a useful mechanism
which can facilitate the escape of a particle from the channel.

\subsection{Optimal resetting rate and resetting transition}\label{opt-r}
The figures (\ref{fig-4}) and (\ref{fig-5}) clearly indicate that for the case whence resetting expedites the search, a unique minima emerges in the mean first passage time. In other words, this states that there is an optimal resetting rate which can minimize the mean time. Formally defining, the optimal rate is given by
\begin{equation}\label{opt-r-eq}
    r^*=\underset{r\ge 0}{\arg \min} \langle T_r(x_0)\rangle,
\end{equation}
where the other parameters are kept fixed.
Using the result (\ref{bw-surv-reset}) with the identity (\ref{mfpt}), the minimization can be performed to evaluate $r^*$ numerically. Generically, the optimal rate will depend crucially on the initial position and the potential depth. Naturally, the question arises: what determines the existence of a finite optimal resetting rate? It was shown only recently that $CV>1$ condition is a sufficient condition for the existence of a finite $r^*>0$. 
In other words, $CV>1$ condition guarantees the existence of atleast one minima and thus a finite optimal resetting rate. In Fig.~\ref{fig-6}(b), we have shown a contour plot spanned in  $(x_0,U_0)$-plane for a symmetric channel superimposed with the optimal resetting rate in color code. It is seen that resetting is helpful for a range of initial positions $x_0$ close to the edges and this range can be further increased by making the potential depth larger. As shown by the inset of  Fig.~\ref{fig-6}(b), at a fixed $U_0$, the optimal rate $r^*$ vanishes at a critical length $x_0^c$ inside the channel which is seen to shift as we vary $U_0$. The transition of $r^*>0$ to $r^*=0$ as we vary $U_0$ is a reminiscent of continuous phase transition and the scaling can be found to be $r^* \sim |x_0-x_0^c|^\alpha$, with $\alpha=1$ as shown from a general theory of resetting transition \cite{pal2019first,ahmad2019first}. Thus, the phase diagram in Fig.~\ref{fig-6}a provides access to the parameters manipulating which we can explore the effects of resetting.

\section{Conditional exit time statistics from the channel}\label{fj-pdf-w-reset}
The shape of the channel clearly suggests that the particle can escape from either of the boundaries. So far, we have looked into unconstrained exits without having any preference to the boundary. However, constrained exits are also important for practical purposes as one maybe interested into the transport through one or a set of specific pores or boundaries. These conditional exits require more detailed information on the nature of trajectories and how they propagate through the channel. In particular, these trajectories intricately depend on the shape of the potential, distance of the exit points from the initial condition and also the resetting rate. In this section we aim to look into the statistics of these conditional exit time statistics and furthermore detail out the effects of resetting.

\begin{figure*}[t]
    \centering
    \includegraphics[scale=0.5]{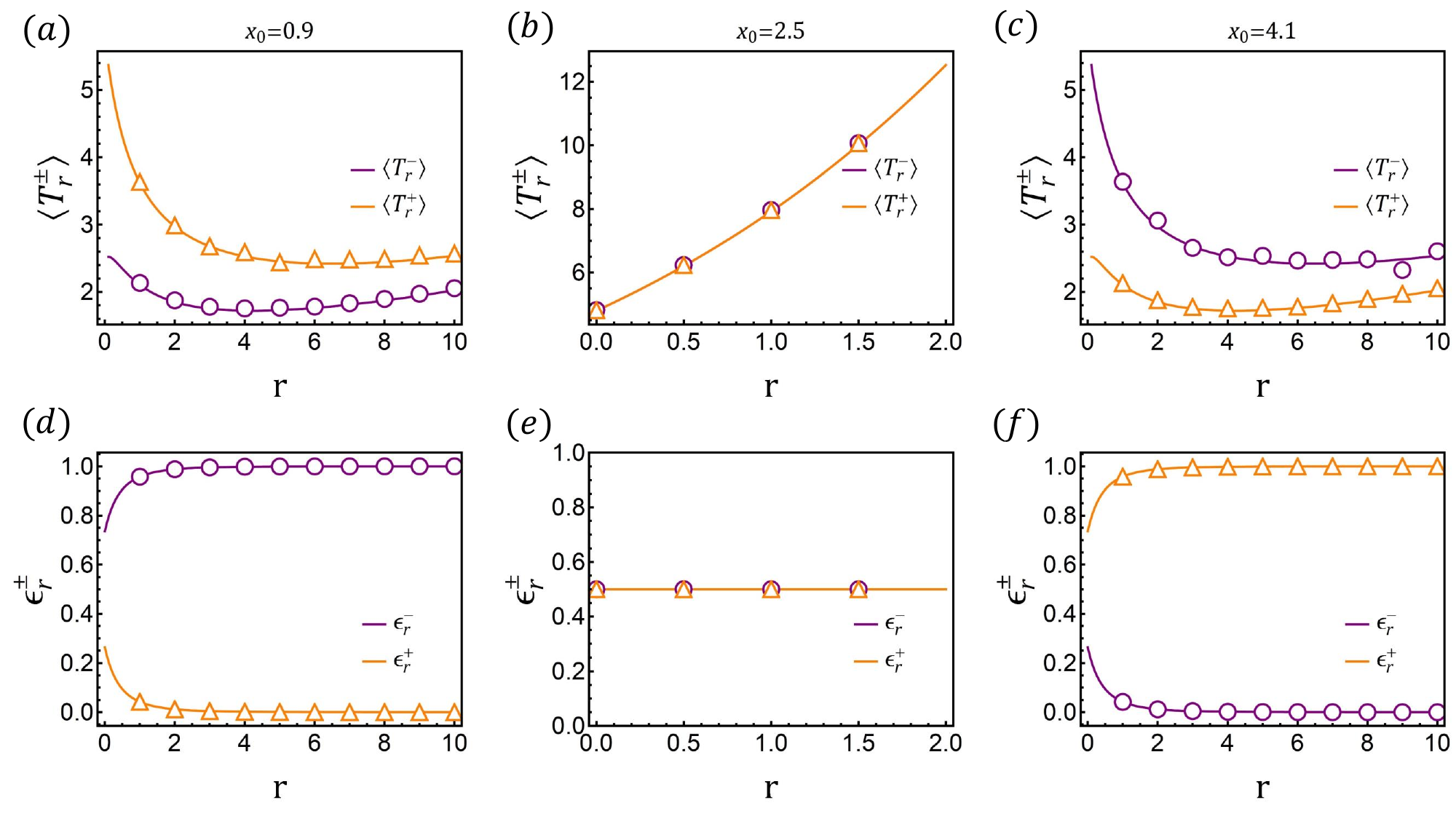}
    \caption{Theoretical and numerical corroboration of conditional MFPT (top) and splitting probabilities (bottom) as a function of the resetting rate. The three columns correspond to three different initial locations, from left to right respectively: $x_0=0.9$, $x_0=2.5$ and $x_0=4.1$, while the rest of the parameters are set fixed $a=1,~b=4,~U_0=1.5,~D=1,~\mu=0.5$, $l=5$. Solid lines describe the theoretical results (using Eqs.~\eqref{eps-r-p},\eqref{eps-r-m}~\&~\eqref{cond-fpt}) whereas the markers indicate simulation results. We observe that the conditional mean first passage times can be minimized with the resetting for the asymmetric initial conditions as depicted in panels (a) and (c), whereas resetting turns out to be detrimental for the symmetric initial condition -- panel (b). From panels (d) and (f) it is observed that the resetting becomes useful in escaping from the left and right end respectively, with its maximum efficiency. In other words, resetting can be tuned to enhance/depreciate the escape probabilities in asymmetric conditions. However, resetting has no impact on the escape probabilities when the particle starts from the center of the channel as depicted in panel (e).}
    \label{fig-7}
\end{figure*}

Unlike the unconditional exits, here one needs to track the trajectories exiting from respective boundaries. To quantify these, one would need the probability density of the particle under resetting dynamics which we define as $p_r(x,t|x_0,0)$. The renewal structure of the resetting dynamics allows us to write an equation for the probability density in the following way \cite{evans2011diffusion,evans2020stochastic}
\begin{align}\label{last-renew-prob}
    p_r\left(x,t|x_0,0\right)&=e^{-rt}p_0\left(x,t|x_0,0\right)+
    r\int_{0}^{t} d\tau e^{-r\tau}p_0\left(x,\tau|x_0,0\right)Q_r\left(t-\tau|x_0\right),
\end{align}
where the first term in Eq.~\eqref{last-renew-prob} corresponds to those trajectories that did not undergo a single resetting event and the second term accumulates the contributions from those trajectories which undergo multiple resetting events until time $t$ without exiting the channel.
The probability of resetting is indicated by the factor $r$ while the condition of not exiting the channel upto time $t$ in the presence of resetting is captured by the survival probability $Q_r(t|x_0)$ that was introduced in the previous section.
Taking Laplace transformation on both sides of Eq.~\eqref{last-renew-prob} and combining with Eq.~\eqref{last-renew-lap}, we find
\begin{equation}\label{last-renew-prob-lap}
    \widetilde{p}_r(x,s|x_0,0)=\dfrac{\widetilde{p}_0(x,r+s|x_0,0)}{1-r\widetilde{Q}_0(r+s|x_0)},
\end{equation}
which connects the probability density functions for the process with and without resetting respectively. To track a specific set of trajectories exiting through a specific boundary, we need to compute the current or particle flux through each of the boundaries. Within our set-up, we denote the particle flux as $J_r^{\pm}(x_0,t)$ where $\pm$ indicates the right or left boundary and they are defined as $J_r^{\pm}(x_0,t)=\mp D \left.\dfrac{\partial p_r(x,t)}{\partial x}\right|_{x=l\ {\rm or}\ 0}$. The exit or splitting probabilities that the particle escapes through a specific boundary is easy to compute once the flux is known namely
\begin{align}
    \epsilon^\pm_r(x_0)=\int_{0}^{\infty} dt~J^{\pm}_r(x_0,t),
\end{align}
where $\epsilon^+_r(x_0)$ and $\epsilon^-_r(x_0)$ are the probabilities that a particle starting at $x_0$ exits to the right and to left respectively, with the normalization $\epsilon^+_r(x_0)+ \epsilon^-_r(x_0)=1$. The conditional exit time through a specific boundary can be estimated by measuring the escape time of the particles through that boundary 
suitably normalized by the particle flux through that same exit point. Formally, they can be written as
\begin{equation}\label{definiton-condi-mfpt}
    \langle T^{\pm}_r(x_0)\rangle=\dfrac{\int_{0}^{\infty} dt~t~J^{\pm}_r(x_0,t)}{\int_{0}^{\infty} dt~J^{\pm}_r(x_0,t)}.
\end{equation}
Although the time dependent quantities such as $J^{\pm}_r(x_0,t)$ are difficult to compute directly, in Laplace space they take simpler form and furthermore can be connected to the probability density via Eq. (\ref{last-renew-prob-lap}). This simple procedure allows us to compute the observables exactly. For instance, the exit probabilities can be rewritten as 
\begin{eqnarray}
\epsilon^+_r(x_0)&&=\widetilde{J}_r^+(x_0,s\rightarrow 0)=-D\dfrac{\partial_x \widetilde{p}_0(x,r|x_0,0)|_{x=l}}{1-r\widetilde{Q}_0(r|x_0)},\label{eps-r-p}\\
\epsilon^-_r(x_0)&&=\widetilde{J}_r^-(x_0,s\rightarrow 0)=D\dfrac{\partial_x \widetilde{p}_0(x,r|x_0,0)|_{x=0}}{1-r\widetilde{Q}_0(r|x_0)},\label{eps-r-m}
\end{eqnarray}
while the mean escape times can be reformulated in the following way from Eq. (\ref{definiton-condi-mfpt})
\begin{equation}\label{cond-fpt}
    \langle T^{\pm}_r(x_0)\rangle=\dfrac{-\left.\dfrac{\partial\widetilde{J}^{\pm}_r(x_0,s)}{\partial s}\right|_{s\rightarrow 0}}{\widetilde{J}^{\pm}_r(x_0,s\rightarrow 0)}=-\left.\dfrac{\partial}{\partial s}\ln\left[\widetilde{J}^{\pm}(x_0,s) \right]\right|_{s=0},
\end{equation}
where one can again use the relation between the current and the underlying probability density function in the Laplace space. Thus, the exit probabilities as a function of resetting rate are obtained by combining Eqs. (\ref{sols-wo-res}), (\ref{bw-surv}) with (\ref{eps-r-p})-(\ref{eps-r-m}), where $\widetilde{p}_0(x,s)$ are given in Appendix \ref{fw-approach} and the conditional MFPTs follow similarly from Eq.~\eqref{cond-fpt}. 
In Fig.~\ref{fig-7}, we plot these results as a function of the resetting rate. We consider the effect of the initial position $x_0$, for a symmetric potential that partially covers a channel of length $l=5$ over a distance $b-a=3$.

To gather statistics for conditional times, we need to track two different sets of trajectories linked with each boundary for a given initial condition. For the symmetric landscape of potential, the favorable escapes depend crucially on the initial condition $x_0$. If the particles start at the proximity to the left boundary, then the escape probability of the particle from the left channel is higher than the other end. The converse effect can be obtained for those $x_0$, closer to the right end. Effectively, one may expect that $\epsilon^-_0$ decreases to $0$ from $1$ as $x_0$ varies from $0\rightarrow l$, whereas $\epsilon^+_0$ increases conversely for the same variation of $x_0$. When the particle starts its journey from $x_0=l/2$ the escape events are identical \textit{i.e.}, $\epsilon^-_0(x_0=l/2)=\epsilon^+_0(x_0=l/2)=\dfrac{1}{2}$. Conditional escape times should also be prioritized accordingly.

\begin{figure*}[t]
    \centering
    \includegraphics[scale=0.6]{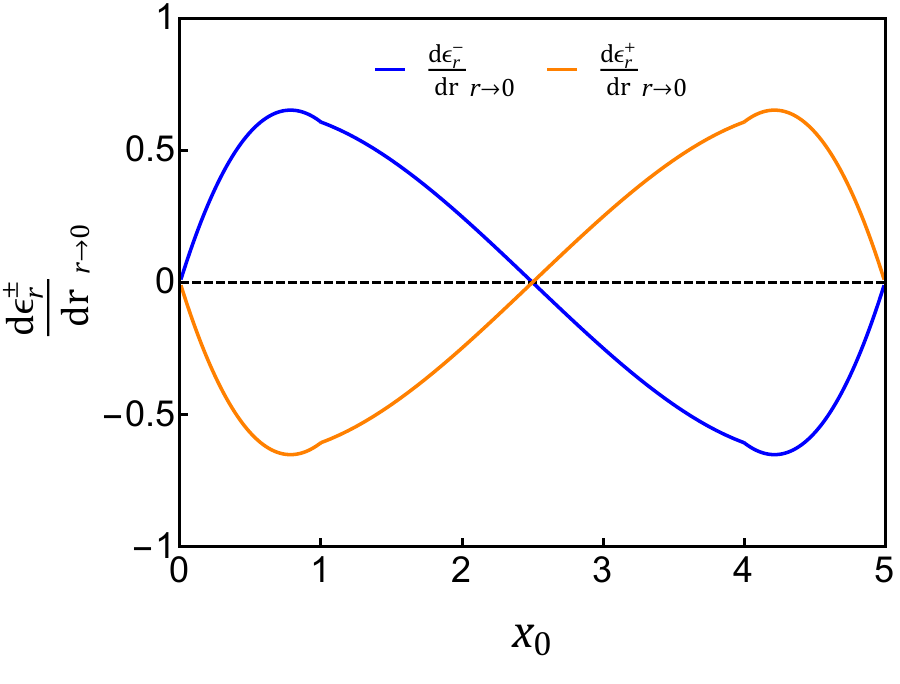}
    \caption{The rate of change in the escape probabilities associated with the left, right boundary as a function of $x_0$, which are represented with the blue and orange solid lines respectively. For instance, when the particle starts from  $x_0<l/2$, the slope is always positive for $\epsilon_r^-$ (blue curve) as we expect the particles to escape from the left boundary more frequently. Eventually, $\epsilon_r^-$ becomes unity as we increase the resetting rate (see also Fig. \ref{fig-7}d). As we switch the initial conditions to $x_0>l/2$, the particle has more probability to escape from the right boundary and furthermore, with resetting $\epsilon_r^+$ is enhanced to unity (see Fig. \ref{fig-7}f). This is evident from the positive value of the slope for $\epsilon_r^+$ for $x_0>l/2$ (orange curve). The slope vanishes exactly at $x_0=l/2$ which is the midpoint of the channel and thus the escape is symmetric with probabilities one-half (also see Fig. \ref{fig-7}e). Here, the parameters were set at: $a=1,~b=4,~U_0=1.5,~D=1,~\mu=0.5,~l=5$. }
    \label{fig-8}
\end{figure*}

Additionally, imposing resetting on the dynamics causes optimization in conditional mean escape times for some initial conditions. From Figs.~\ref{fig-7}(a)~\&~(c), we observe that resetting minimizes the mean conditional escape times for asymmetric initial conditions $x_0=0.9~\&~4.1$. Moreover, the trajectories associated with the long excursions can be discarded with intermittent returns in the presence of resetting; as a consequence, the biasing of escapes becomes stronger (see Figs.~\ref{fig-7}(d)~\&~(f)). For higher resetting rates, these probabilities saturate to maximum value which is unity. In addition, we observe in Fig.~\ref{fig-7}(b), that resetting prolongs the escapes for the symmetric initial condition $x_0=2.5=l/2$, and the escape probabilities can not be optimized further. To understand better, we further analyze the behavior of the rate of escape probabilities i.e., $d\epsilon_{\pm}/dr$ in the limit of $r\rightarrow 0$. As the initial condition is varied from the left side to the right side of the channel, we observe a systematic optimal behavior from the escape trajectories barring the symmetric initial conditions (see Fig.~\ref{fig-8}). For all $x_0$, like in the case of a free particle diffusing in an interval ($U_0=0$) \cite{pal2019first}, resetting tends to favour the escape through one end at the expense of the other, as shown by the fact that $\epsilon^+_r(x_0)\gg \epsilon^-_r(x_0)$ or $\epsilon^+_r(x_0)\ll \epsilon^-_r(x_0)$ at large $r$. Nevertheless, when the particle starts at the bottom of the potential ($x_0=2$), the difference between the splitting probabilities becomes less pronounced. In summary, we conclude that the initial condition plays a crucial role in the resetting mediated transport statistics.

\section{conclusion}\label{conclusion}
In summary, we have investigated the escape statistics of a Brownian particle confined in a channel and subject to a binding potential in the presence of resetting dynamics. Attractive interactions tend to hold the particle for a longer time inside the channel thus prolonging the escape time. Thus, designing navigational strategies which can mitigate such delays are of much need. Here, we show that the resetting dynamics can eliminate looping as well as wandering trajectories inside the channel and can render the transport more efficient. In effect, resetting reduces the mean and the fluctuations of the escape times drastically and we have underpinned the physical scenarios behind such effects. Moreover, we have sketched out a phase diagram in the parameter space spanned by the initial position and the potential strength that provides us a priori knowledge on the modulation of the parameters that could be useful for harnessing the escape from the channel. 

We approached the problem using the Fick-Jacobs formalism that allows one to reduce the three-dimensional diffusion problem to one dimension with an effective diffusivity containing the details of the channel geometry. After studying in detail the unconditional exit time and its modulation with respect to the resetting rate, we looked into the optimal behavior of the escape times and how they can be quantified in practice. 

Overall, this study indicates that resetting can accelerate the crossing or escape time of particles, a characteristic shared by numerous other stochastic processes that involve resetting. However, this effect is not random; it has been consistently shown that resetting is advantageous when stochastic fluctuations in the completion time of a random process are significant, specifically when the coefficient of variation in the absence of resetting ($CV$) exceeds unity \cite{pal2023random,pal2022inspection}. We have explicitly verified this criterion for various potential wells. The effect of resetting has been further analyzed by observing the behavior of the optimal resetting rate that minimizes the mean escape time. In addition, we have reported a continuous transition for the optimal resetting rate when the initial/resetting position is varied.
Extending this formalism, we computed the escape time statistics conditioned on the escape nature of the particles through different boundaries. The conditional escape times are found to display nontrivial behavior as the resetting rate is varied. It
is noticed that resetting can induce a faster escape for the
trajectories which start very close to an undesired boundary and escape through the further boundary. Looping
transient trajectories inside the potential well, which usually cause large fluctuations in such cases, get curtailed by resetting hence a
speed-up in the overall first passage time is observed.

The problem of matter transport through confined
geometries containing narrow openings, corners and bottlenecks
has recently led to many theoretical and experimental research fronts. Recent single particle experiments have also shown interesting observations namely identifying the structure of a channel or generically the number of intermediate configurations from the first passage or escape time statistics of various living and non-living agents \cite{thorneywork2020direct}. It would be worthwhile to look into this inference problem within our set-up eg. whether one can infer the number of bottlenecks from the first passage time statistics from a channel. It is evident that resetting frequency can be used as a probe to mitigate particle clogging inside the channel which can otherwise hinder the transport -- it will be interesting to investigate such effects in soft-matter experiments who deal with colloidal particles and resetting \cite{tal2020experimental,goerlich2023experimental,besga2020optimal,faisant2021optimal}.

\section{Acknowledgements}
AP gratefully acknowledges the DST-SERB Start-up Research Grant Number SRG/2022/000080 and the Department of Atomic Energy, India for research funding (via the project `Modeling of Soft Materials'). This study was partially supported by CONACyT under the grant Frontiers Science No. 51476. The authors thank the Raman Research Institute, Bangalore for the hospitality during the conference `Frontiers in Statistical Physics' where some initial discussion took place. The authors also thank the Higgs Center for Theoretical Physics, Edinburgh for hospitality during the workshop `New Vistas in Stochastic resetting' where the project concluded.

\appendix

\section{Probability density functions without resetting}\label{fw-approach} 
To utilize the renewal relation in Eq.~\eqref{last-renew-prob}, it is essential to know the propagator of the underlying reset-free process in Laplace space. In this section, we solve the forward Smoluchowski equation for the probability density function, $\widetilde{p}_0\left(x,s|x_0,0\right)$ with absorbing boundaries at two ends of the channel.
Taking Laplace transformation of Eq.~\eqref{pdf}, we get
\begin{align}\label{fw-pdf-lap}
    s \widetilde{p}_0\left(x,s\right)-&\delta \left(x-x_0\right)=
    D\dfrac{\partial}{\partial x}\left[e^{-\beta U(x)}\dfrac{\partial}{\partial x}\left(e^{\beta U(x)}\widetilde{p}_0\left(x,s \right)\right)\right].
\end{align}
As mentioned earlier, the shape of the potential allows us to solve Eq.~\eqref{fw-pdf-lap} in a piece-wise manner in three different regions. Regime wise i.e., based on the final position, we denoted these corresponding propagators as $\widetilde{p}_0^{L,M,R}(x,s|x_0,s)$ in the main text. However, the particle can propagate to any one of these regimes at the final time $t$ starting either from the same regime or elsewhere at time $0$. Thus, it is meaningful also to split these probability densities based on the exact initial condition. The notation for the break-up in the initial condition is denoted by $I,II,III$. For instance, $\widetilde{p}_0^{L,I}(x,s|x_0,0)$ denotes the probability density function of the particle to be found in the domain $0< x< a$ given that it had started in the interval $0< x_0< a$ at time zero. Likewise, other propagators can also be written based on the final position and the starting point. In what follows, we solve Eq.~\eqref{fw-pdf-lap} in detail based on these splittings.

\subsection{Interval $0< x_0< a$}\label{regime-1}
We first consider that the initial position $x_0$ of the particle lies to the left part of the channel (referred to by the index $I$ in the expression below), but the final position $x$ can be in any of the three regions. Thus, Eq.~\eqref{fw-pdf-lap} can be modified as
\begin{equation}\label{left-pdf}
    \begin{cases}
        s\widetilde{p}_0^{L,I}\left(x,s\right)-\delta\left(x-x_0\right)=D\dfrac{\partial^2}{\partial x^2}\widetilde{p}_0^{L,I}\left(x,s\right)~~~~~0< x< a,\\
        \\
        s\widetilde{p}_0^{M,I}\left(x,s\right)=D\dfrac{\partial^2}{\partial x^2}\widetilde{p}_0^{M,I}\left(x,s\right)~~~~~a< x< b,\\
        \\
        s\widetilde{p}_0^{R,I}\left(x,s\right)=D\dfrac{\partial^2}{\partial x^2}\widetilde{p}_0^{R,I}\left(x,s\right)~~~~~b< x< l,\\
    \end{cases}
\end{equation}
where we denote $\widetilde{p}_0^{L,I},\widetilde{p}_0^{M,I}$ and $\widetilde{p}_0^{R,I}$ as solutions of the probability density function in the Laplace domain, and the suffixes denote the left, middle, and right solutions respectively, for $x_0$ belonging to the left part. Using the absorbing boundary conditions, one can solve Eqs.~\eqref{left-pdf} and the corresponding solutions read
\begin{equation}\label{sol-left-pdf}
    \begin{cases}
     \widetilde{p}_0^{L,I}\left(x,s|x_0,0\right)=\dfrac{1}{2 \alpha_0 D}e^{-\alpha (x+x_0)}\left[2 \alpha_0 D A^{L,I} \left(e^{2 \alpha  x}-1\right) e^{\alpha  x_0}+\Theta (x-x_0) \left(e^{2 \alpha  x_0}-e^{2 \alpha  x}\right)\right],\\
     \\
     \widetilde{p}_0^{M,I}\left(x,s|x_0,0\right)=A^{M1,I}e^{\alpha_0 x}+A^{M2,I}e^{-\alpha_0 x},\\
     \\
     \widetilde{p}_0^{R,I}\left(x,s|x_0,0\right)=A^{R,I}e^{-\alpha_0 x}\left(e^{2\alpha_0 x}-e^{2\alpha_0 l}\right),
    \end{cases}
\end{equation}
where while solving the first equation in Eq. \ref{sol-left-pdf} we have also used the following matching conditions due to the Dirac delta function: (i) continuity of the probabilities, (ii) discontinuity in the derivative of the probabilities across the delta function.

One can derive the relations among the constants $A^{L,I}$, $A^{M1,I}$, $A^{M2,,I}$ and $A^{R,I}$ in Eqs.~\eqref{sol-left-pdf} by applying
the matching conditions~\eqref{cont-bc}-\eqref{jump-bc}:
\begin{equation}\label{cons-left-pdf}
    \begin{cases}
        2\sinh{\left[a \alpha_0\right]}A^{L,I}=\dfrac{1}{\alpha_0 D}\sinh{\left[(a-x_0)\alpha_0\right]}+e^{-a \alpha_0-U_0\beta}\left(A^{M1,I}e^{2a \alpha_0}+A^{M2,I}\right),\\
        \\
        A^{M1,I}e^{2b \alpha_0}+A^{M2,I}=A^{R,I}e^{U_0\beta}\left(e^{2b\alpha_0}-e^{2l\alpha_0}\right),\\
        \\
        \alpha_0 D e^{-a\alpha_0}\left[A^{L,I}+e^{2a\alpha_0}\left(A^{L,I}-A^{M1,I}\right)+A^{M2,I}\right]=\cosh{\left[(a-x_0)\alpha_0\right]},\\
        \\
        \alpha_0 e^{b\alpha_0} A^{M1,I}=\alpha_0 e^{-b\alpha_0}\left[A^{M2,I}+A^{R,I}\left(e^{2b\alpha_0}+e^{2l\alpha_0}\right)\right].
    \end{cases}
\end{equation}
These coefficients can be obtained by using Mathematica.
One can gather all the solutions given in Eqs.~\eqref{sol-left-pdf} in one equation as
    \begin{equation}\label{fw-left-pdf}
        \widetilde{p}_0^I\left(x,s|x_0,0\right)=\Theta\left(a-x\right)\widetilde{p}_0^{L,I}\left(x,s|x_0,0\right)+\Theta\left(x-a\right)\Theta\left(b-x\right)\widetilde{p}_0^{M,I}\left(x,s|x_0,0\right)+\Theta\left(x-b\right)\widetilde{p}_0^{R,I}\left(x,s|x_0,0\right),
    \end{equation}

\subsection{Interval $a< x_0< b$}\label{regime-2}
Similarly, we rewrite Eq.~\eqref{fw-pdf-lap} when the initial position lies in the middle part of the potential (referred to by the index $II$ below), 
\begin{equation}\label{middle-pdf}
    \begin{cases}
        s\widetilde{p}_0^{L,II}\left(x,s\right)=D\dfrac{\partial^2}{\partial x^2}\widetilde{p}_0^{L,II}\left(x,s\right)~~~~~0< x< a,\\
        \\
        s\widetilde{p}_0^{M,II}\left(x,s\right)-\delta\left(x-x_0\right)=D\dfrac{\partial^2}{\partial x^2}\widetilde{p}_0^{M,II}\left(x,s\right)~~~~~a< x< b,\\
        \\
        s\widetilde{p}_0^{R,II}\left(x,s\right)=D\dfrac{\partial^2}{\partial x^2}\widetilde{p}_0^{R,II}\left(x,s\right)~~~~~b< x< l.\\
    \end{cases}
\end{equation}
From the boundary conditions, the solutions of Eqs.~\eqref{middle-pdf} read
\begin{equation}\label{sol-middle-pdf}
    \begin{cases}
        \widetilde{p}_0^{L,II}\left(x,s|x_0,0\right)=2A^{L,II}\sinh{\left[\alpha_0 x\right]},\\
        \\
        \widetilde{p}_0^{M,II}\left(x,s|x_0,0\right)=A^{M1,II}e^{\alpha_0 x}+A^{M2,II}e^{-\alpha_0 x}-\dfrac{1}{\alpha_0 D}\sinh{\left[(x-x_0)\alpha_0\right]}\Theta\left(x-x_0\right),\\
        \\
        \widetilde{p}_0^{R,II}\left(x,s|z,0\right)=-2A^{R,II}e^{L\alpha_0}\sinh{\left[(l-x)\alpha_0\right]}.
    \end{cases}
\end{equation}
Also, the relations among the constants are derived as in Sec.~\ref{regime-1} from the matching conditions \eqref{cont-bc}-\eqref{jump-bc},
\begin{equation}\label{cons-middle-pdf}
    \begin{cases}
        2\sinh{\left[a\alpha_0\right]}A^{L,II}=e^{-\beta U_0}\left(e^{a\alpha_0}A^{M1,II}+e^{-a\alpha_0}A^{M2,II}\right),\\
        \\
        -2A^{R,II}e^{L\alpha_0}\sinh{\left[(l-b)\alpha_0\right]}=
        e^{-\beta U_0}\left(A^{M1,II}e^{b\alpha_0}+A^{M2,II}e^{-b\alpha_0}-\dfrac{1}{\alpha_0 D}\sinh{\left[(b-x_0)\alpha_0\right]}\right),\\
        \\
        2A^{L,II}\cosh{\left[a\alpha_0\right]}+A^{M2,II}e^{-a\alpha_0}=A^{M1,II}e^{a\alpha_0},\\
        \\
        e^{b\alpha_0}\alpha_0A^{M1,II}=\dfrac{1}{D}\cosh{\left[(b-x_0)\alpha_0\right]}+\alpha_0 e^{-b\alpha_0} \left[A^{R,II}\left(e^{2l\alpha_0}+e^{2b\alpha_0}\right)+A^{M2,II}\right].
    \end{cases}
\end{equation}

After solving this system of equations with Mathematica, we get the following form of the probability density function in Laplace space

    \begin{equation}\label{fw-middle-pdf}
    \widetilde{p}_0^{II}\left(x,s|x_0,0\right)=\Theta\left(a-x\right)\widetilde{p}_0^{L,II}\left(x,s|x_0,0\right)+\Theta\left(x-a\right)\Theta\left(b-x\right)\widetilde{p}_0^{M,II}\left(x,s|x_0,0\right)+\Theta\left(x-b\right)\widetilde{p}_0^{R,II}\left(x,s|x_0,0\right).
\end{equation}

\subsection{Interval $b< x_0< l$}\label{regime-3}
When the initial condition lies in the last, rightmost segment (referred to as by the index $III$), Eq.~\eqref{fw-pdf-lap} is written as
\begin{equation}\label{right-pdf}
    \begin{cases}
        s\widetilde{p}_0^{L,III}\left(x,s\right)=D\dfrac{\partial^2}{\partial x^2}\widetilde{p}_0^{L,III}\left(x,s\right)~~~~~0< x< a,\\
        \\
        s\widetilde{p}_0^{M,III}\left(x,s\right)=D\dfrac{\partial^2}{\partial x^2}\widetilde{p}_0^{M,III}\left(x,s\right)~~~~~a< x< b,\\
        \\
        s\widetilde{p}_0^{R,III}\left(x,s\right)-\delta\left(x-x_0\right)=D\dfrac{\partial^2}{\partial x^2}\widetilde{p}_0^{R,III}\left(x,s\right)~~~~~b< x< l.\\
    \end{cases}
\end{equation}
From the boundary conditions, we get the following form for the solutions of Eq.~\eqref{right-pdf}
\begin{equation}\label{sol-right-pdf}
    \begin{cases}
        \widetilde{p}_0^{L,III}\left(x,s|x_0,0\right)=2A^{L,III}\sinh{\left[\alpha_0 x\right]},\\
        \\
        \widetilde{p}_0^{M,III}\left(x,s|x_0,0\right)=A^{M1,III}e^{\alpha_0 x}+A^{M2,III}e^{-\alpha_0 x},\\
        \\
        \widetilde{p}_0^{R,III}\left(x,s|x_0,0\right)=\\
        \\
        \dfrac{1}{2D\alpha_0}e^{-(x+x_0)\alpha_0}\left[e^{2l\alpha_0}-e^{2x_0\alpha_0}+\left(e^{2x_0\alpha_0}-e^{2b\alpha_0}\right)\Theta\left(x-x_0\right)\right]\\
        \\
        ~~~~~~~-A^{R,III}e^{x_0\alpha_0}\left(e^{2l\alpha_0}-e^{2x\alpha_0}\right).
    \end{cases}
\end{equation}
Using the matching conditions \eqref{cont-bc}-\eqref{jump-bc},
the constants $A^{L,III}$, $A^{M1,III}$, $A^{M2,III}$ and $A^{R,III}$ satisfy the relations
\begin{equation}\label{cons-right-pdf}
    \begin{cases}
        2\sinh{\left[a\alpha_0\right]}A^{L,III}=e^{-\beta U_0}\left(e^{a\alpha_0}A^{M1,III}+e^{-a\alpha_0}A^{M2,III}\right),\\
        \\
        \dfrac{e^{-(b+x_0)\alpha_0}}{2D\alpha_0}\left[e^{2l\alpha_0}-e^{2x_0\alpha_0}+2\alpha_0 D e^{x_0\alpha_0}\left(e^{2b\alpha_0}-e^{2l\alpha_0}\right)A^{R,III}\right]\\
        \\
        ~~~~~=e^{-\beta U_0}\left(A^{M1,III}e^{b\alpha_0}+A^{M2,III}e^{-b\alpha_0}\right),\\
        \\
        2\cosh{\left[a\alpha_0\right]}A^{L,III}+e^{-a\alpha_0}A^{M2,III}=e^{a\alpha_0}A^{M1,III},\\
        \\
        \alpha_0\left(e^{b\alpha_0}A^{M1,III}-e^{-b\alpha_0}A^{M2,III}\right)=2\alpha_0 e^{b\alpha_0}A^{R,III}\\
        \\
        +\dfrac{1}{2D}e^{-(b+x_0)\alpha_0}\left[e^{2x_0\alpha_0}-e^{2l\alpha_0}-2De^{x_0\alpha_0}\left(e^{2b\alpha_0}-e^{2l\alpha_0}\right)\alpha_0A^{R,III}\right].
    \end{cases}
\end{equation}
After solving this system with Mathematica, one gathers the solutions in a single form as

\begin{equation}\label{fw-right-pdf}
    \widetilde{p}_0^{III}\left(x,s|x_0,0\right)=\Theta\left(a-x\right)\widetilde{p}_0^{L,III}\left(x,s|x_0,0\right)+\Theta\left(x-a\right)\Theta\left(b-x\right)\widetilde{p}_0^{M,III}\left(x,s|x_0,0\right)+\Theta\left(x-b\right)\widetilde{p}_0^{R,III}\left(x,s|x_0,0\right).
\end{equation}

Combining all the probability density functions from Eqs.~\eqref{fw-left-pdf},\eqref{fw-middle-pdf} \& \eqref{fw-right-pdf} in one equation, one can write a full solution for the probability density function in the following way

\begin{equation}\label{fw-total-pdf}
    \widetilde{p}_0\left(x,s|x_0,0\right)=\Theta\left(a-x_0\right)\widetilde{p}_0^{I}\left(x,s|x_0,0\right)+\Theta\left(x_0-a\right)\Theta\left(b-x_0\right)\widetilde{p}_0^{II}\left(x,s|x_0,0\right)+\Theta\left(x_0-b\right)\widetilde{p}_0^{III}\left(x,s|x_0,0\right).
\end{equation}

\section{Derivation of the matching conditions in Eqs.~\eqref{imperfect-bound} and \eqref{flux-cont}}\label{appendix-b}
The survival probability $Q_0(t|x_0)$ of the Brownian particle without resetting inside the $1d$ channel satisfies the backward Fokker-Planck equation
\begin{equation}\label{b1}
    \dfrac{\partial Q_0}{\partial t}=D\dfrac{\partial^2 Q_0}{\partial x_0^2}-\mu U'(x_0)\dfrac{\partial Q_0}{\partial x_0},
\end{equation}
where $D$ is the diffusion constant, $\mu$ the mobility and $U(x)$ the external potential. This equation can be rewritten as 
\begin{eqnarray}\label{b2}
    \dfrac{\partial Q_0}{\partial t}=De^{\beta U(x_0)}\dfrac{\partial}{\partial x_0}\left(e^{-\beta U(x_0)}\dfrac{\partial Q_0}{\partial x_0}\right),
\end{eqnarray}
where $\beta=\mu/D=(k_BT)^{-1}$. Using the Laplace transform of $Q$,
\begin{equation}\label{b3}
    \widetilde{Q}_0(s|x_0)=\int_0^\infty dt~e^{-st}Q_0(t|x_0),
\end{equation}
the transform of Eq.~\eqref{b2} gives
\begin{equation}\label{b4}
    -1+s\widetilde{Q}_0=De^{\beta U(x_0)}\dfrac{\partial}{\partial x_0}\left(e^{-\beta U(x_0)}\dfrac{\partial \widetilde{Q}_0}{\partial x_0}\right),
\end{equation}
where the initial condition $Q_0(t=0|x_0)=1$ for any position $x_0$ other than the absorbing boundaries has been applied.

To determine a first matching condition at $x=a$ given the discontinuous potential in Eq. (\ref{potential-form}), let us multiply Eq.~\eqref{b4} by $e^{-\beta U(x_0)}$ and integrate in a small interval $\left[a-\epsilon,a+\epsilon\right]$ around $a$,
\begin{equation}\label{b5}
    \int_{a-\epsilon}^{a+\epsilon}dx_0~e^{-\beta U(x_0)}\left[-1+s\widetilde{Q}_0(s|x_0)\right]=D\left[e^{-\beta U(x_0)}\dfrac{\partial \widetilde{Q}_0}{\partial x_0}\right]_{a-\epsilon}^{a+\epsilon}.
\end{equation}
The LHS of Eq.~\eqref{b5} tends to $0$ as $\epsilon\rightarrow 0$, because $\widetilde{Q}_0(s|x_0)$ does not have any singularity (since $0\leq Q_0(t|x_0)\leq 1$ then $0\leq\widetilde{Q}_0(s|x_0)\leq 1/s$). One deduces the boundary condition
\begin{equation}
    \dfrac{\partial \widetilde{Q}_0(s|a^+)}{\partial x_0}=e^{-\beta U_0}\dfrac{\partial \widetilde{Q}_0(s|a^-)}{\partial x_0},\label{b6}
\end{equation}
and similarly,
\begin{equation}
    \dfrac{\partial \widetilde{Q}_0(s|b^+)}{\partial x_0}=e^{\beta U_0}\dfrac{\partial \widetilde{Q}_0(s|b^-)}{\partial x_0}.\label{b7}
\end{equation}
Since Eqs.~\eqref{b6}~and~\eqref{b7} are valid for any $s$, they also hold in the time domain.

To derive the other matching condition at $x_0=a$, we take the Laplace transform of Eq.~\eqref{b1} and use the identity $U'(x)=-U_0\delta(x-a)$ for the discontinuous potential in the left part of the channel. We thus have
\begin{equation}\label{b8}
    -1+s\widetilde{Q}_0=D\dfrac{\partial^2 \widetilde{Q}_0}{\partial x_0^2}+\mu U_0\delta(x_0-a)\dfrac{\partial\widetilde{Q}_0}{\partial x_0}.
\end{equation}
To avoid to deal with the discontinuity of $\partial \Tilde{Q}_0/\partial x_0$ at $a$ during integration, let us multiply the above equation by $x_0-a$. Making the change of variable $u=x_0-a$,
\begin{equation}\label{b9}
    u\left[-1+s\widetilde{Q}_0(s|u)\right]=Du\dfrac{\partial^2 \widetilde{Q}_0(s|u)}{\partial u^2}+\mu u U_0\delta(u)\dfrac{\partial\widetilde{Q}_0(s|u)}{\partial u}.
\end{equation}
We again integrate over $u$ in a small interval $[-\epsilon,\epsilon]$. The first term of the rhs of Eq.~\eqref{b9} can be integrated by parts and the second one is identically $0$. Hence,
\begin{equation}\label{b10}
    \int_{-\epsilon}^{\epsilon} du~u~\left[-1+s\widetilde{Q}_0(s|u)\right]=D\epsilon \left(\dfrac{\partial\widetilde{Q}_0(s|\epsilon)}{\partial u}+\dfrac{\partial\widetilde{Q}_0(s|-\epsilon)}{\partial u}\right)-D\left[\widetilde{Q}_0(s|\epsilon)-\widetilde{Q}_0(s|-\epsilon)\right].
\end{equation}
We see that the lhs and the first term of the rhs of Eq.~\eqref{b10} tend to $0$ as $\epsilon\rightarrow 0$, therefore we conclude that $\widetilde{Q}_0$ is continuous at $x_0=a$,
\begin{equation}\label{b11}
    \widetilde{Q}_0(s|a^+)=\widetilde{Q}_0(s|a^-).
\end{equation}
Similarly near $x_0=b$,
\begin{equation}\label{b12}
    \widetilde{Q}_0(s|b^+)=\widetilde{Q}_0(s|b^-).
\end{equation}
These relations clearly hold in the time domain as well. The matching conditions derived here are crucial to compute the survival probability through the entire channel. 

\section{Relations among the coefficients of Eq.~\eqref{sols-wo-res} for $\widetilde{Q}_0(s|x_0)$}\label{relation-coeff}
Here, we provide the necessary relations to evaluate the coefficients $\{A,B,C_1,C_2\}$ in Eq.~\eqref{sols-wo-res}. Using the matching conditions \eqref{imperfect-bound}-\eqref{flux-cont} we get the following relations
\begin{equation}\label{coeff-surv}
    \begin{cases}
        \dfrac{e^{-\alpha_0 a}}{s}+2A\cosh{\alpha_0 a}=e^{\mu U_0/D}\left(C_1e^{\alpha_0 a}-C_2e^{-\alpha_0 a}\right),\\
        e^{\alpha_0 l}+s~B\left(e^{2\alpha_0 b}+e^{2\alpha_0 l}\right)=s~e^{\mu U_0/D}\left(e^{\alpha_0 b}C_1-C_2\right),\\
        s~C_2=s~(A-C_1)e^{2 \alpha_0 a}-A~s-1,\\
        s~C_2=s~(B-C_1)e^{2\alpha_0 b}-e^{\alpha_0 l}-s~Be^{2\alpha_0 l}.
    \end{cases}
\end{equation}
These coefficients are obtained explicitly with Mathematica. Since their expressions are rather long, we do not reproduce them here.

\section{Details of the numerical simulations and implementation of the boundary conditions}\label{simul}
\begin{figure}
    \centering
    \includegraphics[width=\columnwidth]{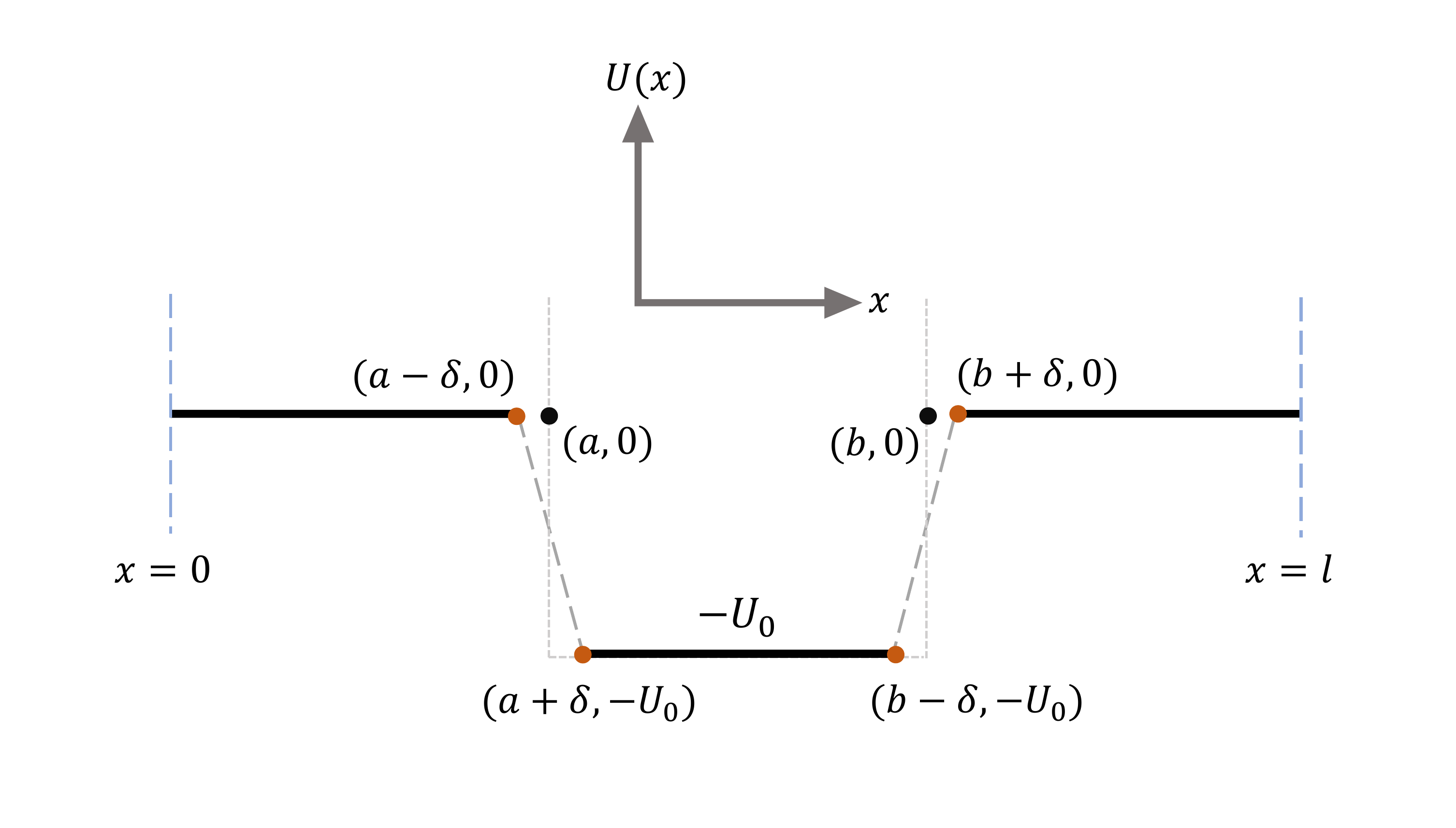}
    \caption{Schematical representation of the form of the potential in terms of two linear potentials.}
    \label{fig-9}
\end{figure}
In this section, we elaborate on the simulation method of our model. The presence of discontinuities in the potential implies delta functions in the force. Here, we replace the discontinuities of the box-shaped potential given in Eq.~\eqref{potential-form} by two linear functions with a large slope,
\begin{equation}
    U(x)=
    \begin{cases}
        0 & 0<x<a-\delta\\
        -U_0\dfrac{x-(a-\delta)}{2\delta} & a-\delta\leq x\leq a+\delta\\
        -U_0 & a+\delta<x<b-\delta\\
        U_0\dfrac{x-(b+\delta)}{2\delta} & b-\delta\leq x\leq b+\delta\\
        0 & b+\delta<x<l,
    \end{cases}
\end{equation}
where $\delta$ is the standard deviation of 
the free Brownian displacement between two consecutive times $t$ and $t+\Delta t$. Hence, $\delta=\sqrt{2 D~\Delta t}$ with $\Delta t$ the numerical time step of the simulation. We run the Langevin dynamics with the force field given by $f(x)=-\dfrac{d}{dx}U(x)$. Starting from $x_0$ at $t=0$, we have updated the particle position in the following way:
\begin{enumerate}
    \item Call a random number $\lambda$ from the uniform distribution in $(0,1)$ and check the resetting condition: if $\lambda<r \Delta t$ then bring the particle back to the initial position $x_0$, otherwise perform step 2.
    \item Update the position as
    \begin{equation}\label{langevin}
        x(t+\Delta t)=x(t)+\mu f(x(t))\Delta t+\sqrt{2D~\Delta t}~\xi(t+\Delta t),
    \end{equation}
    where $\xi(t+\Delta t)$ is a normally distributed random number with zero mean and unit variance. As usual, $\mu$ is the mobility of the particle and $D$ is the diffusion coefficient.
    \item Depending on the updated and previous positions of the particle, we invoke the force term. The force acts on the particle when $x(t)\in(a-\delta,a+\delta)\cup(b-\delta,b+\delta)$. When the particle crosses the point $x=a$, it might also happen that $|x(t+\Delta t)-x(t)|>2\delta$ with $|x(t)-a|>\delta$, i.e., the particle did not feel the force term. To overcome this difficulty, we have added the constant force term in Eq. (\ref{langevin}) despite of having $|x(t)-a|>\delta$, when the Brownian term of Eq. (\ref{langevin}) alone makes the particle cross the force field near the position $x=a$. 
    \item To track the first passage time, we impose the boundary conditions: \textit{i.e.}, $x(t+\Delta t)<0\bigcup x(t+\Delta t)>l$. If one of these conditions is satisfied, we terminate the simulation, make a note of the corresponding time step number $n$, multiply with the microscopic time step $\Delta t$ and the total time $n\Delta t$ contributes to the unconditional first passage time. Over many realizations, we collect these random times and gather the first passage time statistics. 
\end{enumerate}

\nocite{*}
\bibliography{final}

\end{document}